\documentstyle[12pt]{article}
\begin{document}
\begin{titlepage}
\hbox to \hsize{\hfil  Preprint IHEP 97-26}
\hbox to \hsize{\hfil  April 1997}
\hbox to \hsize{\hfil  hep-ph/9704228}
\hbox to \hsize{\hfil }
\hbox to \hsize{\hfil }
\hbox to \hsize{\hfil }
\vfill
\large \bf
\begin{center}
Analyticity and Minimality 
of Nonperturbative Contributions in Perturbative Region 
for $\bar \alpha_s$
\end{center}
\vskip 1cm
\normalsize
\begin{center}
{\bf Aleksey I. Alekseev
\footnote{Electronic address: 
alekseev@mx.ihep.su} 
and
Boris A. Arbuzov
\footnote{Electronic address: 
arbuzov@mx.ihep.su}
}\\
{\small \it Institute for High Energy Physics,\\
  Protvino, Moscow Region, 142284 Russia}
\end{center}
\vskip 1.5cm
\begin{abstract}
It is shown, that the possibility of a freezing 
of QCD running coupling constant at zero in the 
approach with "forced analyticity" can not be in accord with 
Schwinger-Dyson equation for gluon propagator. 
We propose to
add to the analytic expression the well-known 
infrared singular term $1/q^2$ as well as pole term corresponding to  
"excited gluon". With this example we formulate the principle of 
minimality of nonperturbative contributions in perturbative 
(ultraviolet) region, which allows us to fix ambiguities in 
introduction of nonperturbative terms and maintain the finiteness 
of the gluon condensate.  
As a result we obtain estimates of the gluon condensate, which quite 
agree with existing data. The nonzero effective mass of the 
"excited gluon" leads also to some interesting qualitative 
consequences.
\end{abstract}
\vskip 1cm
PACS number(s): 12.38.Aw, 12.38.Lg
\vfill
\end{titlepage}

\newcommand{\bi}{\bibitem}
\newcommand{\be}{\begin{equation}}
\newcommand{\ee}{\end{equation}}

The discovery of the asymptotic freedom property~\cite{Gros} in
non-Abelian gauge theories turned to be a decisive factor in
the formation QCD as the strong interaction theory. The negative 
sign of QCD $\beta$-function
$\beta (g^2)=\beta_0g^4+..., \beta_0 =-b_0/(16\pi^2),
b_0=11 C_2/3-2N_f/3$ in the vicinity of zero provided the number of
active quarks being not too large
(for $SU_c(3) \;\; N_f\leq 16$) gives coupling constant,
which describes quarks and gluons interaction at large  Euclidean 
$q^2$, i.e. at small distances,
\begin{equation}
\bar g^2(q^2/\mu^2, g)=\frac{g^2}{1-\beta_0 g^2\ln (q^2/\mu^2)},
\label{a1}
\end{equation}
tending towards zero. Therefore in the deep Euclidean region we are
allowed to use perturbation theory. In expression~(\ref{a1}), which
takes into account the main logarithms, $\mu$ is a normalization 
point.
An account of the next $g^2$ corrections does not change asymptotic
behaviour~(\ref{a1}) for $q^2\to\infty$. By introducing dimensional
constant $\Lambda^2=\mu^2\exp (-4\pi/(b_0\alpha_s)),
\alpha_s=g^2/4\pi$, we turn from explicitly renormalization
invariant expression ~(\ref{a1}) to the following formula
\begin{equation}
\bar\alpha_s (q^2)=\frac{4\pi}{b_0\ln (q^2/\Lambda^2)}.\label{a2}
\end{equation}
It is reasonable to estimate parameter $\Lambda$ in
approximate expression~(\ref{a2}) to be around of few hundreds MeV. 
With decreasing  $q^2$ effective constant~(\ref{a2}) increases,
that may indicate a tendency of unlimited growth of the interaction 
at large distances, leading to a confinement of coloured objects.
However, at $q^2=\Lambda^2$ in expression~(\ref{a2}) the pole is 
present, 
which is nonphysical at least due to failing of the perturbation 
theory,
starting from which formula~(\ref{a2}) has been obtained.

In recent work~\cite{Shir} a solution of the problem of ghost pole
was proposed with imposing of a condition of analyticity in $q^2$.
The idea of "forced analyticity" goes back to works ~\cite{Red,Bog}
of the late fifties, which were dedicated to the problem of
Landau-Pomeranchuk pole~\cite{Land} in QED. Using for $\bar\alpha_s(q^2)$
a spectral representation without subtractions, the following 
expression
for the running coupling constant was obtained in paper~\cite{Shir}
\be
\bar\alpha_s^{(1)}(q^2)=\frac{4\pi}{b_0}\left [
\frac{1}{\ln(q^2/\Lambda^2)} +\frac{\Lambda^2}{\Lambda^2-q^2}\right ].
\label{a3}
\ee
This expression has the asymptotic freedom property and its
analyticity in the infrared region is due to nonperturbative
contributions. It does not contain any additional parameter and 
has finite limit at zero, $\bar\alpha_s^{(1)}(0)=4\pi/b_0\simeq 
1.40$ (freezing of the coupling constant), 
which depends only on symmetry factors. This limit turns 
out to be stable with respect to higher orders corrections.

As it is noted in work~\cite{Bog}, a procedure of summation of 
leading 
logarithmic terms is not defined uniquely. A partial fixation of 
this ambiguity in QED is realized by using of a method of 
summation of the perturbation theory series under the sign of the 
spectral integral of the K\"allen-Lehmann representation. 
Nevertheless after such summation 
a functional ambiguity remains, which from one side does not violate 
correct analytic properties of Green functions 
in a complex plane of a corresponding invariant variable and from 
the other side contains nonanalytic dependence on constant $g^2$. 
In work~\cite{Arb} while investigating the photon propagator in QED
it was shown, that ambiguities in summation procedure of the 
diagram series can be 
removed provided one demands not only the validity of spectral 
representation, but also the fulfillment of equations of motion. 

In the present note we consider a problem of consistency of the
constant behaviour of the effective charge in the infrared region 
with Schwinger-Dyson (SD) equation for a gluon 
propagator. Further we include into consideration nonperturbative
terms, the singular
in the infrared region term $\sim 1/q^2$ in particular, the necessity 
of the renormalization invariance being taken into account. Then we 
discuss
possibilities of an adjustment of demands of confinement, asymptotic 
freedom,
analyticity, an accordance with the perturbation theory and 
correspondence with estimates of the gluon condensate value.

To study the problem of a possibility of a constant behaviour of the 
running
constant in the infrared region let us consider the integral 
SD  equation for the gluon propagator in ghost-free 
axial gauge~\cite{Kum}
$A^a_\mu\eta_\mu=0,\;\; \eta_\mu$ --- gauge vector, $\eta^2\not=0$.
In this gauge the effective charge is directly connected with the
gluon propagator and Slavnov-Taylor identities \cite{Slav} have
the most simple form. 
The important preference of the axial gauge consist in the
possibility to exclude the term from SD equation, which
contains the full four-gluon vertex by means of contraction
of the equation with tensor $\eta_\mu\eta_\nu/\eta^2$.

In what follows we shall work in the Euclidean momentum space,
where smallness of the momentum squared is immediately connected
with smallness of its components. The equation to be considered
has the form:
$$
[D^{-1}_{\mu\nu}(p)-D^{-1}_{(0)\mu\nu}(p)]\frac{\eta_\mu\eta_\nu}
{\eta^2}=\Pi_{\mu\nu}(p)
\frac{\eta_\mu\eta_\nu}{\eta^2}, 
$$
$$
\Pi_{\mu\nu}(p)=-\frac{C_2g^2\mu^{4-n}}{2(2\pi)^n}\int d^nk
\Gamma^{(0)}_{3\mu\lambda\rho}(p,-k,k-p)D_{\lambda\sigma}(k)
D_{\rho\delta}(p-k)\times
$$
\be
\times\Gamma_{3\sigma\delta\nu}(k,p-k,-p),\label{a29}
\ee
where $\Pi_{\mu\nu}(p)$ is the one-loop part of the polarization 
operator, 
$D_{\mu\nu}(p)$ is the propagator, 
$\Gamma_{\sigma\delta\nu}(k,p-k,-p)$ is the 
one-particle irreducible three-gluon vertex function, 
$\Gamma^{(0)}_{\mu\lambda\rho}(p,-k,k-p)$ is the free three-gluon 
vertex. 

We suppose the approximation $D_{\mu\nu}(p) = Z(p^2) D_{(0)\mu\nu}
(p)$ to be appropriate to study the infrared region. 
Let us divide the momentum integration domain in expression
~(\ref{a29}) in two parts:
$k^2<\lambda^2$ and $k^2>\lambda^2$, where $\lambda$ is sufficiently
small, but finite. Then domain $k^2>\lambda^2$ in the case of absence 
of kinematic singularities 
in three-gluon vertex
gives a contribution, 
which is regular in $p^2$ for $p^2\to 0$, and in domain 
$k^2<\lambda^2$
full Green functions can be approximated by free ones up to constant
factors according to an assumption of running constant be frozen at 
zero. 
Then one can write
$$
\Pi_{\mu\nu}(p)\frac{\eta_\mu\eta_\nu}{\eta^2}=-
\frac{C_2g^2\mu^{4-n}Z(0)}{2(2\pi)^n}\int\limits^{\lambda}_{0}
d^nk\Gamma^{(0)}_{3\mu\lambda\rho}(p,-k,k-p)\times
$$
\be
\times D^{(0)}_{\lambda\sigma}(k) D^{(0)}_{\rho\delta}(p-k)
\Gamma^{(0)}_{3\sigma\delta\nu}(k,p-k,-p)
\eta_\mu\eta_\nu/\eta^2+
Q(p^2;y,\lambda,n).\label{a30}
\ee
Here $y\,=\,(p\eta)^2/p^2 \eta^2$ is the gauge parameter. 
The integration in formula~(\ref{a30}) can be extended up to all
the domain of momentum, that results in a change of
the regular in $p^2$ contribution $Q$. Thus one has
\be
\Pi_{\mu\nu}(p)\frac{\eta_\mu\eta_\nu}{\eta^2}=Z(0)\Pi^{(1)}_{\mu\nu}
(p)\frac{\eta_\mu\eta_\nu}{\eta^2}+Q(p^2;y,n), \label{a31}
\ee
where $\Pi^{(1)}_{\mu\nu}(p)$ is the one-loop 
perturbation theory contribution to the polarization operator. 
This contribution is
calculated in paper~\cite{A_ICTP} and has rather complicated
structure. Let us present the expression for the leading terms
of convolution~(\ref{a31}) at $y\to 0$. We have
$$
\Pi^{(1)}_{\mu\nu}(p)\frac{\eta_\mu\eta_\nu}{\eta^2}=
Cp^2\left [-\frac{22}{3\epsilon}-\frac{22}{3}\left(\gamma-2+\ln
\frac{p^2}{4\pi\mu^2}\right)-
\frac{70}{9}+\right.
$$
\be
\left.+\frac{40}{3}y\ln y+O(y,y^2\ln y)\right ]. \label{a32}
\ee
Here $C=g^2C_2/32\pi^2$, $\gamma$ is the Euler constant.
From expression~(\ref{a32}) we see, that singularity at $y=0$
is smooth and the limit at $y=0$ does exist.
Term $\sim 1/\epsilon$ $(n=4+2\epsilon)$ as well as constant ones
could be absorbed into function $Q$ while the logarithm of the
momentum squared necessarily persists. The equation for function
$Z(p^2)$ takes the form
\be
Z^{-1}(p^2)=1+Z(0)\frac{g^2C_2}{16\pi^2}\frac{11}{3}\ln p^2+Q(p^2;n).
\label{a33}
\ee
We see, that behaviour $Z(p^2)\simeq Z(0)\not=0$ for $p^2\to 0$
does not agree with the SD equation. 

This conclusion stimulate us to look for 
the possibilities different from the assumption  on the finiteness 
of the coupling constant at zero. Recently the possibility of the 
soft singular power infrared behaviour of the gluon propagator was 
discussed~\cite{Cud}, $D(q) \sim (q^2)^{-\beta}$, $q^2 \rightarrow 0$, 
where $\beta$ is a small positive non-integer number. In 
Ref.~\cite{AlekPL} the consistency of such behaviour
with Eq.~(\ref{a29}) was studied. A characteristic equation for the 
exponent $\beta$ was obtained and this equation was shown not to 
have solutions in the region $0 < \beta < 1$. The authors of 
Ref.~\cite{Butner} also come to the conclusion on the inconsistency 
of the soft singular infrared behaviour of
the gluon propagator. The case of possible interference of power 
terms was studied in Ref.~\cite{AkekTMF96} and it was shown that in 
a rather wide interval $-1 < \beta < 3$ of the non-integer values 
of the exponent the characteristic equation has no solutions. 
At present
the more singular, in comparison with free case, infrared behaviour 
of the form $D(q) \simeq M^2/(q^2)^{2}$, $q^2 \to 0$ seems to be the
most justified~\cite{Pagels,Baker,Alek1}. 
The physical consequences of such enhancement of zero modes are 
discussed in the reviews~\cite{ArbPartNucl,RobWil}. 
Bearing in mind the remarks stated above let us consider the 
following expression for the running coupling:
\be
\bar\alpha_s(q^2)=\frac{4\pi}{b_0}\left [
\frac{1}{\ln q^2/\Lambda^2}+\frac{\Lambda^2}{\Lambda^2-q^2}+c
\frac{\Lambda^2}{q^2}\right ].\label{a35}
\ee
Let us represent this expression in explicitly renormalization 
invariant form. It can be done without solving the differential 
renormalization group equations. In this order we write
$\bar\alpha_s(q^2)=\bar g^2(q^2/\mu^2,\; g^2)/4\pi$ and use the 
normalization condition $\bar g^2(1,g^2)=g^2$. Then we obtain
the equation for wanted dependence of the parameter $\Lambda^2$ on
$g^2$ and $\mu^2$:
$$
g^2/4\pi=\frac{4\pi}{b_0}
\left [
\frac{1}{\ln \mu^2/\Lambda^2}+\frac{\Lambda^2}{\Lambda^2-\mu^2}+c
\frac{\Lambda^2}{\mu^2}\right ].
$$
From dimensional reasons
$
\Lambda^2=\mu^2 exp\{-\varphi (x)\}$,
where $x=b_0g^2/16\pi^2=b_0\alpha_s/4\pi$, and for function
$\varphi (x)$ we obtain the equation:
$$
x=\frac{1}{\varphi (x)}+\frac{1}{1-e^{\varphi (x)}}+ce^{-\varphi 
(x)}.
$$
The solution of this equation at $c>0$  is monotone decreasing 
function $\varphi (x)$, which has the behaviour $\varphi (x)\simeq 
1/x$ at $x\to 0$ and  $\varphi (x)\simeq -\ln(x/c)$ at
$x\to +\infty$. 
The relation obtained ensures the renormalization invariance of
$\bar\alpha_s(q^2)$. At low  $g^2$ we obtain
$
\Lambda^2=\mu^2\exp\{-4\pi/(b_0\alpha_s)\}$,
which indicates the essentially nonperturbative character of
both last terms of the Eq.~(\ref{a35}) and this terms are absent
in the perturbation theory. With given value of the QCD scale 
parameter $\Lambda$ the parameter
$c$ can be fixed by the string tension $\kappa$ or the Regge slope 
$\alpha'=1/(2\pi \kappa)$ assuming the linear confinement
$V(r)\simeq \kappa r=a^2r$ at $r\to \infty$. 
We define the potential $V(r)$ of static $q\bar q$ interaction
~\cite{BoShir,Buch} by means of three-dimensional Fourier 
transform of $\bar \alpha_s(\vec q^2)/\vec q^2$ with 
the contributions of only one dressed gluon exchange taken 
into account. This gives the following relation
\be
c\Lambda^2=(3b_0/8\pi)a^2=(b_0/16\pi^2)g^2M^2. \label{aa36}
\ee
At large $q^2$ from Eq.~(\ref{a35}) one obtains
\be
\bar\alpha_s(q^2)=\frac{4\pi}{b_0}\left [
\frac{1}{\ln q^2/\Lambda^2}+(c-1)\frac{\Lambda^2}{q^2}-
\frac{\Lambda^4}{(q^2)^2}+ O((q^2)^{-3})\right ].\label{a41}
\ee
From Eq.~(\ref{a41}) it is seen that in the ultraviolet region
the nonperturbative contributions decrease more rapid then 
all renormalization group improved perturbation theory corrections. 
The value  $c =1$ corresponds to maximal suppression of 
nonperturbative contributions in the ultraviolet region. Accepting 
this condition one obtains the connection of the QCD
scale parameter  $\Lambda$  and string tension  $\kappa = a^2$
of the form
$\Lambda^2= 3b_0 \kappa / 8\pi\,$. 
Taking  $a\simeq 0.42\,GeV$ one obtains for $\Lambda$ 
reasonable estimation, $\Lambda \simeq 0.434\, GeV$ ($b_0 = 9$ in
the case of 3 light flavours). 

Considering the nonperturbative contributions the following 
arguments can be expressed. One knows QCD to be renormalizable 
in the perturbation theory and, as usually, the renormalization 
procedure can be developed to remove the divergences in all orders.
However, what about the nonperturbative contributions?
If they bring in the additional divergences then the problem of
renormalization turns out to be unsolved. The situation when
nonperturbative contributions do not violate the perturbative
renormalization properties seems to be more attractive.
It take place if the nonperturbative contributions decrease
at momentum infinity sufficiently fast and  do not introduce the 
divergences in observables. So, it is natural to demand their 
fastest of possible decrease at large momenta. 
An application of the principle
of minimality of nonperturbative contributions in the ultraviolet
region  will be shown further with taking as
an example the important physical quantity, namely, the gluon
condensate, 
$K\,=\,{< \alpha_s/\pi:G^a_{\mu \nu}\,G^a_{\mu \nu}:>}\,$. 
According to the definition (see e.g.,~\cite{ArbPartNucl}) 
up to the quadratic approximation in the gluon fields one has
after Wick rotation
\be
K\,=\,\frac{48}{\pi}\,\int \frac{d^4k}{(2 \pi)^4}\,
\left(\bar\alpha_s(k^2)\, - \,\bar\alpha_s^{pert}(k^2)\right)\,=\,
\frac{3}{ \pi^3} \,\int_0^{\infty}\,\bar\alpha_s^{np}(y)
\,y dy\, ,\label{a44}
\ee
where $\bar\alpha_s^{np}$ is nonperturbative part of the 
running coupling constant. 
In our case the two last terms of Eq.~(\ref{a35}) should be taken.
By substituting this terms in Eq.~(\ref{a44}) one can see the 
logarithmic divergences of the integral at infinity and at finite 
point $k^2=\Lambda^2$.

The acceptance of the cancellation mechanism for the nonphysical
perturbation theory singularities~(\ref{a2}) by the nonperturbative 
contributions leads to the necessity of supplementary definition of
the integral~(\ref{a44}) near point  $k^2 = \Lambda^2$.
This problem can be reformulated as a problem of dividing
of perturbative and nonperturbative contributions in $\bar\alpha_s$
resulting in introduction of some parameter $k_0=1\div2$ GeV.
This provides absence of the pole at $k^2=\Lambda^2$ in 
both perturbative and nonperturbative parts.
The divergence of the integral~(\ref{a44})
at infinity stimulate  the further modification of the running 
coupling
constant.
Going over from Eq.~(\ref{a3}) to Eq.~(\ref{a35}) the isolated 
singularity was introduced. In this case the singularity 
corresponding to the unitary cut was not changed and in accordance 
with the approach of Refs.~\cite{Red,Bog,Shir}
is determined by perturbation theory. Following to this logic let us 
consider the expression for $\bar\alpha_s$ with one more isolated 
singularity in the time-like region. The tachion singularity 
in the space-like region, of cause, is prohibited.

The principle of minimality of nonperturbative contributions
in ultraviolet region then leads to the following unique expression
for the running coupling constant
\be      
\bar\alpha_s(q^2)\,=\,\frac{4 \pi}{b_0}\Biggl(\frac{1}{
ln(q^2/\Lambda^2)}\,
+\,\frac{\Lambda^2}{\Lambda^2 - q^2}\,+\,\frac{c \Lambda^2}{q^2}\,+
\,\frac{(1-c) \Lambda^2}{q^2 + m_g^2}\Biggr)\,,\label{a45}
\ee
with fixed residue and mass parameter $m_g$,
\be
m_g^2\,=\,\Lambda^2/(c-1),\label{a46}
\ee
for the newly introduced term.
The expression~(\ref{a45}) can be represented in explicitly 
renormalization invariant form in a similar way to the 
expression~(\ref{a35}).
Nonperturbative contributions in Eq.~(\ref{a45}) decrease at 
infinity as $1/q^6$, the integral in Eq.~(\ref{a44}) converges and 
we can obtain
\be
K\,=\,\frac{4}{3 \pi^2}\,\Lambda^4\,\{ln(c-1) +
k^2_0/\Lambda^2 + ln(k^2_0/\Lambda^2 - 1)\}
 \,.\label{a47}
\ee

Phenomenology gives the positive value of the gluon condensate $K$ 
in the interval  $(0.32\,GeV)^4$ -- 
$(0.38\,GeV)^4$~\cite{Vain,Grein}. 
As an example we take values $k_0=1.2 \div 1.3$ GeV.
If one regards the string tension parameter to be given, then from 
Eqs.~(\ref{a46}), (\ref{a47}) and~(\ref{aa36})
one has the dependencies of all the values under consideration
on the parameter $c$, which are presented in the Table I.

Note that values  $c = 1.063,\,\Lambda = 422\, MeV,\,
m_g = 1.682\, GeV$, $k_0 = 1.265$ GeV
corresponds to the conventional value of the 
gluon condensate~\cite{Vain}
$K = (0.33\,GeV)^4$. 
Certainly these results should be considered as tentative, 
but nevertheless they seems encouraging.

\begin{table}
\caption{Parameters of the running 
coupling constant~(13) 
and gluon condensate as functions of parameter $c$.}
\begin{center}
\begin{tabular}{c c c c c c} \hline
$c$  & $\Lambda$, GeV & $m_g$, GeV & $K^{1/4}$, GeV & $K^{1/4}$, GeV
 & $K^{1/4}$, GeV \\ 
     &       &       & $k_0=1.2$ GeV & $k_0=1.25$ GeV & $k_0=1.3$ GeV
 \\ \hline
1.01 & 0.433 & 4.332 & 0.298 & 0.309 & 0.318 \\ 
1.02 & 0.431 & 3.048 & 0.307 & 0.317 & 0.326 \\ 
1.03 & 0.429 & 2.476 & 0.312 & 0.321 & 0.330  \\ 
1.04 & 0.427 & 2.134 & 0.315 & 0.324 & 0.332  \\ 
1.05 & 0.425 & 1.900 & 0.317 & 0.326 & 0.334  \\ 
1.06 & 0.423 & 1.726 & 0.319 & 0.327 & 0.335  \\ 
1.07 & 0.421 & 1.591 & 0.320 & 0.328 & 0.336  \\ 
1.08 & 0.419 & 1.481 & 0.321 & 0.329 & 0.337  \\ 
1.10 & 0.415 & 1.313 & 0.322 & 0.330 & 0.337  \\ 
1.12 & 0.411 & 1.187 & 0.323 & 0.330 & 0.337  \\ 
1.16 & 0.404 & 1.010 & 0.323 & 0.330 & 0.337 \\ 
1.20 & 0.397 & 0.889 & 0.322 & 0.329 & 0.336 \\ 
1.24 & 0.391 & 0.798 & 0.321 & 0.328 & 0.335  \\ 
1.30 & 0.382 & 0.697 & 0.319 & 0.326 & 0.332  \\ \hline
\end{tabular}
\end{center}
\end{table}
It is seen from Eq.~(\ref{a45}) that the pole singularities are
situated at two points $q^2 = 0$ and $q^2 = - m_g^2$.
It corresponds to two effective gluon masses, $0$ and $\,m_g$. 
Therefore the physical meaning of the parameter $m_g$ is not
the constituent gluon mass but rather the mass of the exited state 
of the gluon. It is essential that the residue at $m^2_g$ 
is very small, so the states with the exited gluons should be
quite narrow in contrast to the spectrum of the coupled 
massless gluons.

The qualitative picture of the glueball states corresponding 
to the running coupling constant~(\ref{a45}) 
with $m_g \simeq 1.7$ GeV
could be the following:

1) The states $g\,g$ --- continuous spectrum and very wide 
resonances are probable;

2) The states $g\,g'$ --- resonances with probable mass interval
 1500 -- 1800 MeV  and with width suppression factor $(1 - c)$;

3) The narrow states  $g'\,g'$ --- resonances with possible
 masses 3000 -- 3600 MeV and with width suppression factor 
 $(1 - c)^2$.

Note that in the region 2) there are the glueball candidates.
The region 3) is insufficiently investigated, some indications in 
favour of the narrow states are showing up (see e.g.,~\cite{Aleev}).
\bigskip

We would like to thank Yu.F.~Pirogov and V.E.~Rochev for interesting 
discussion. A.I.A. is grateful also to C.D.~Roberts, 
J.M.~Namys{\l}ovski, and J.P.~Vary for stimulating discussions.

\end{document}